\documentclass[
    ,final            
  ]
  {aipproc}
\usepackage{amssymb}
\layoutstyle{6x9}

\begin{document}

\title{Measuring the local dark matter density}

\addvspace{-2.5cm}
\author{Silvia Garbari}{
  address={Institute of Theoretical Physics, University of Z\"urich, Z\"urich, Switzerland. }
}
\author{George Lake}{
  address={Institute of Theoretical Physics, University of Z\"urich, Z\"urich, Switzerland. }
}
\author{Justin Read$^{*,}$}{
  address={Department of Physics \& Astronomy, University of Leicester,  Leicester,  United Kingdom.}
}

\begin{abstract}
We examine systematic problems in determining the local matter density from the vertical motion of stars, i.e. the ``Oort limit". Using collisionless simulations and a Monte Carlo Markov Chain technique, we determine the data quality required to detect local dark matter at its expected density.  We find that systematic errors are more important than observational errors and apply our technique to \emph{Hipparcos} data to reassign realistic error bars to the local dark matter density.
\end{abstract}

\maketitle
\begin{figure}
  \includegraphics[height=.2\textheight]{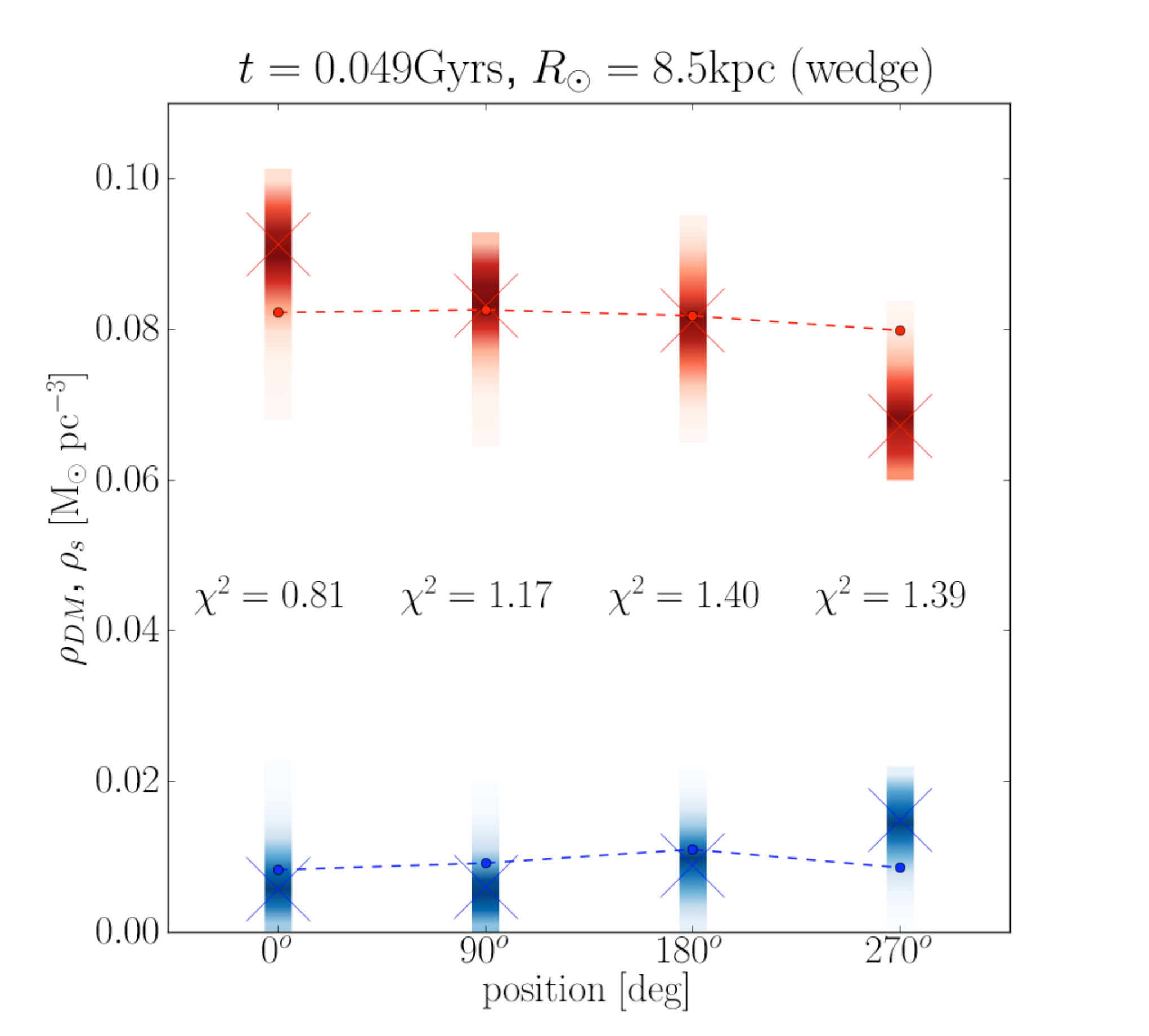}
   \includegraphics[height=.2\textheight]{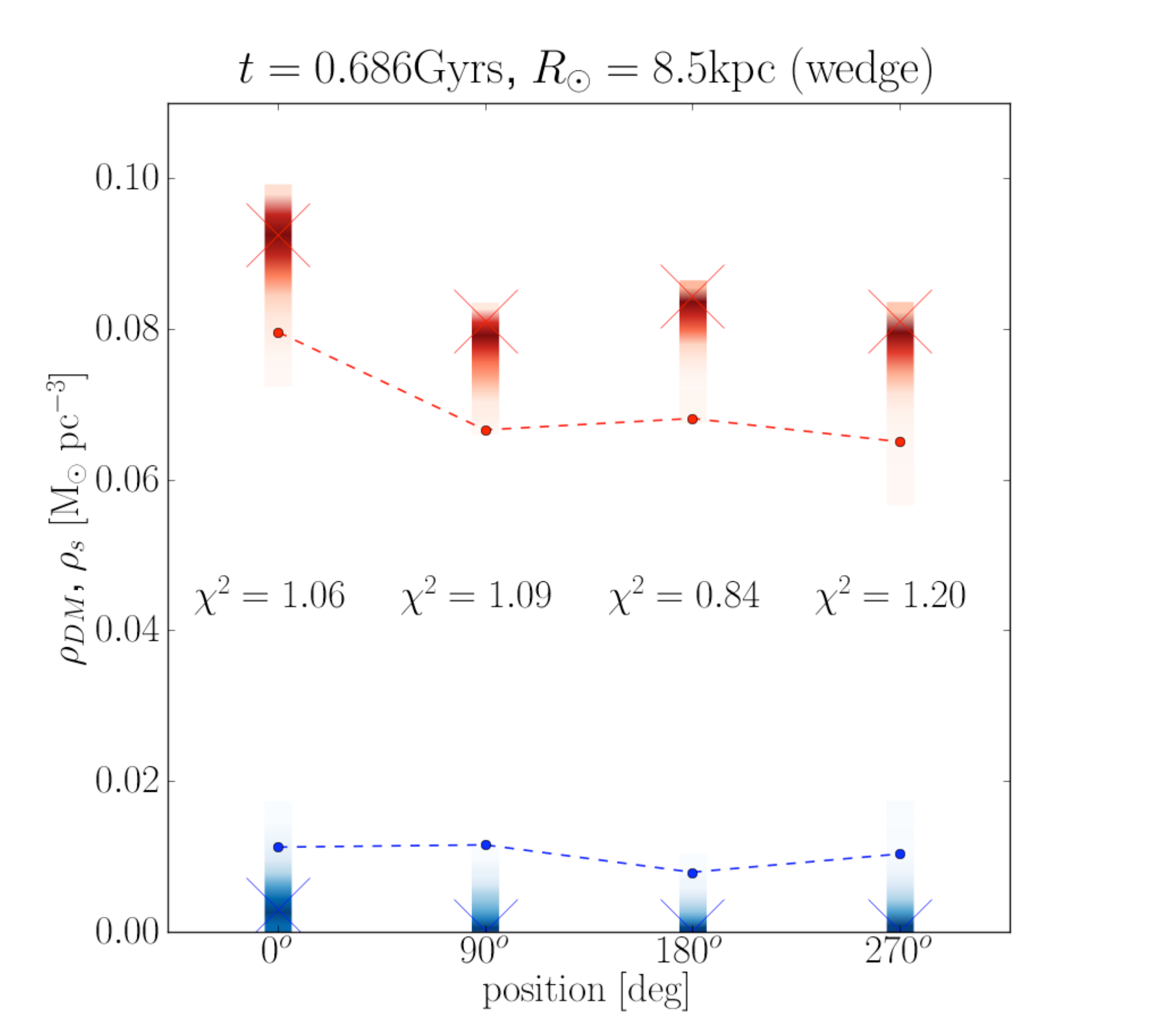}
    \caption{Recovered star (upper rectangles) and DM (lower rectangles) densities for the axisymmetric (left) and inhomogeneous (right) discs in four volumes around the disc. The shaded regions represent the density of the MCMC models and the crosses the best $\chi^2$ values.\vspace*{-.25in}}\label{figura}
\end{figure}

\paragraph{Introduction} Determining the local density of matter is an old problem [2] with a renewed importance
for direct dark matter (DM) searches. The local mass density determined from measurements of stellar motions is  $0.102\pm 0.01$M$_\odot$pc$^{-3}$ 
[3, 4], limited by the sample size of stars with good distance estimates and proper motions.  The quoted uncertainties are of the same order as the DM density in the standard dark halo model for the Galaxy.  Upcoming surveys promise a dramatic 
improvement in the number and precision of astrometric, photometric and spectroscopic measurements.  \\
With this in mind,  we revisit the Oort limit [1].   Following the work of [3, 5, 6], we determine the local DM density by comparing the predicted density  $\nu(z)$ 
found from the vertical velocity distribution at the plane $f(w_0)$ of a stellar tracer,  (equation \ref{den}), 
\vspace*{-0.05in}\begin{equation}
\nu(z)=2\int^{\infty}_{\sqrt{2\Phi}}\frac{f(|w_0|)w_0dw_0}{\sqrt{w_0^2-2\Phi}}
\label{den}
\end{equation}
with the measured stellar density fall-off in cylindrical patch at the Solar Neighborhood.\\
As in [3, 5], we calculate the potential by modeling the local visible mass as a superposition of isothermal components, and the local dark mass as a constant density term.  We then integrate the Poisson and Boltzmann equations simultaneously, neglecting the cross terms of the latter. These approximations are valid as long as the potential can be separated into its vertical and radial component, $\Phi(R,z)=\Phi(z)+\Phi(R)$, the local visible matter is made up of nearly isothermal species and the disk scale height is much smaller than the disk and the halo scale lengths: $z_d\ll r_d \ll R_h$.
\vspace*{-0.1in}\paragraph{Analysis and results} We apply this method to a high resolution (30 million disc particles) collisionless simulation of a MW-like galaxy, whose initial condition were produced using the galaxy models by Widrow and Dubinski [7], and fit the density fall-off with a Monte Carlo Markov Chain technique to find the best value of of the local stellar and DM densities ($\rho_s$ and $\rho_{DM}$). First, we take an early snapshot of the simulation ($\sim$50Myrs) when the disc is still axisymmetric to produce a ``super-patch'' containing a large number of particles at the solar radius, comparable to what we will get with  the next generation surveys (e.g. GAIA). Then we use the evolved simulation ($\sim$0.7Gyrs), which has a bar and spiral waves, to evaluate the impact of disc inhomogeneities.\\
Taking stars in larger volumes around the Sun improves the statistics, but it's important to evaluate how far we can go in $z$ and $R$ before breaking the method's assumptions. From the analysis of the axisymmetric disc, using boxes with different radial and vertical sizes, we can recover the local $\rho_s$ and $\rho_{DM}$ provided we use a local volume to compute the velocity distribution with $|z|\lesssim 75$pc, $R\lesssim 300$pc and we fit the density fall-off up to 3-4 times the disc scale height $z_d$ (i.e. $z>0.75$kpc). This latter is required to break a strong degeneracy between  $\rho_s$ and $\rho_{DM}$ due to the potential similarities at low $z$ .\\
In the axisymmetric disc, we recover the local density correctly within 1$\sigma$ (of the MCMC model distribution) as long as we have a good sampling (e.g. GAIA data) and the systematic errors are of the order of the observational uncertainties and of the expected amount of DM in the standard dark halo model for the Galaxy ($\sim 0.01$M$_\odot$pc$^{-3}$), as shown in figure \ref{figura}.\\
The situation is different in presence of disc inhomogeneities: in this case the errors are larger and the method seems to systematically overestimate the stellar density (figure \ref{figura} right).  This may owe to the cross terms of the velocity dispersion tensor in the inhomogeneous disc, that are neglected in the method.

 The ``super patches'' have statistics that are comparable to the future GAIA sample, so we want to introduce realistic errors in the velocities and positions of the stars to quantify their effects on the recovered density.  We find that these errors don't change the recovered values for the density, but merely increase the $\chi^2$ of the fit. This means that systematic errors have a dominant effect. For this reason, it is important to calculate the local matter density for the available data, using our MCMC technique, to assign realistic errors to the DM density measure.  This will be the next step of our analysis.\\

 \small{\paragraph{Acknowledgements} We would like to thank Larry Widrow and John Dubinski for kindly providing us the software GalactICS2008 to produce the initial conditions and for  the useful suggestions to set up the Galaxy model.}\\

\center{\textbf{REFERENCES}}
\footnotesize{
\begin{itemize}
\item[1.] S.~Garbari, J.~Read and G.~Lake, \emph{in prep.} (2010).
\item[2.] J.~H. Oort, \emph{Bulletin of the Astronomical Institutes of the Netherlands} \textbf{6}, 249 (1932).
\item[3.] J.~Holmberg and C.~Flynn, \emph{MNRAS} \textbf{ 313}, 209--216 (2000).
\item[4.] J.~Holmberg and C.~Flynn, \emph{MNRAS} \textrm{352}, 440 (2004). 
\item[5.] J.~N. Bahcall \emph{et al.}, \emph{ApJ} \textbf{276}, 156 (1984), \emph{ApJ} \textbf{287}, 926 (1984), \emph{ApJ} \textbf{276}, 169 (1984).
\item[6.] B.~Fuchs and C.~Flynn, \emph{MNRAS} \textbf{270}, 471 (1994).
\item[7.] L.~M.~Widrow , J.~Dubinski, \emph{Astrophysical Journal}, \textbf{631}, 838  (2005); L.~M.~Widrow , B.~Pym, J.~Dubinski, \emph{Astrophysical Journal}, \textbf{679}, 1239 (2008).
\end{itemize}}

\end{document}